\documentclass[twocolumn,pre,showpacs]{revtex4}
\usepackage{graphicx}\large
\usepackage{dcolumn}
\usepackage{amsmath}
\begin{document}

\title{Disordered, stretched, and semiflexible biopolymers in two dimensions}
\author{Zicong Zhou}
\email{zzhou@mail.tku.edu.tw} \affiliation{Department of Physics,
Tamkang University, 151 Ying-chuan, Tamsui 25137, Taiwan, Republic
of China}

\author{B\'{e}la Jo\'{o}s}
\email{bjoos@uottawa.ca}
\homepage{http://www.science.uottawa.ca/~bjoos/}
\affiliation{Ottawa Carleton Institute for Physics, University of
Ottawa Campus, Ottawa, Ontario, Canada, K1N-6N5} %\today

\begin{abstract}
We study the effects of intrinsic sequence-dependent curvature for
a two dimensional semiflexible biopolymer with short-range
correlation in intrinsic curvatures. We show exactly that when not
subjected to any external force, such a system is equivalent to a
system with a well-defined intrinsic curvature and a proper
renormalized persistence length. We find the exact expression for
the distribution function of the equivalent system. However, we
show that such an equivalent system does not always exist for the
polymer subjected to an external force. We find that under an external
force, the effect of sequence-disorder depends upon the averaging
order, the degree of disorder, and the experimental conditions,
such as the boundary conditions. Furthermore, a short to moderate
length biopolymer may be much softer or has a smaller apparent
persistent length than what would be expected from the
``equivalent system". Moreover, under a strong stretching force
and for a long biopolymer, the sequence-disorder is immaterial for
elasticity. Finally, the effect of sequence-disorder may depend
upon the quantity considered.

\begin{center}
{\em Accepted to publish in Phys. Rev. E.}
\end{center}

\end{abstract}
 \maketitle

\section{Introduction}
It is known that sequence-dependent properties of biopolymers play
a crucial role in many biological processes. More specifically,
sequence-disorder has important influences on DNA packaging,
transcription, replication, recombination, and repair processes
\cite{TA93,PH88,PN98,BDM98,SFCTFMWW06,PT07,MFFA07,ATVDMA01,VAA05,AVADT02,ZJ08}.
Owing to progress in experimental techniques such as laser or
magnetic tweezers and atomic force microscopy, it is now possible
to manipulate and observe single biomolecules directly, and thus
make a better comparison between theoretical predictions and
experimental observations.

In theoretical studies, a semiflexible biopolymer is often modeled
as a filament \cite{KP49,MS94,BMSS94,MS95,SABBC96,
SFB92,ZJ08,PR00,ZZO00,
ZLJ05,ZJLYJ07,PN98,BDM98,PT07,AVADT02,ATVDMA01,VAA05,MFFA07,ZZ07}.
For instance, the wormlike chain (WLC) model, which views the
biopolymer as an inextensible chain with a uniform bending
rigidity but with a negligible cross section, has been used
successfully to describe the entropic elasticity of a long
double-stranded DNA (dsDNA) \cite{KP49,MS94,BMSS94,MS95}. However,
the traditional elastic models are usually uniform and ignore the
role of sequence-disorder. Under what conditions such a
simplification is valid is therefore an intriguing question. Based
on the elastic models, two effects of sequence-disorder need to be
considered. First, structural inhomogeneity yields variations of
the bending rigidity along the filament, and results in an
$s$-dependent persistence length $l_p(s)$ \cite{PT07,ZJ08}, where
$s$ is the arc length. It has been reported that for a long
biopolymer with short-range correlation (SRC) in $l_p(s)$ and free
of external force, this effect can be accounted by a replacement
of the $l_p(s)$ by an appropriate average \cite{PT07,ZJ08}.
However, for a short biopolymer, inhomogeneity in $l_p(s)$ tends
to make physical observables divergent \cite{PT07,ZJ08}. Second,
the local structure can be characterized by the intrinsic
sequence-dependent curvatures (i.e., the static curvature or the
frozen-in curvature)
\cite{PH88,PN98,BDM98,SFCTFMWW06,PT07,MFFA07,ATVDMA01,VAA05,AVADT02,ZJ08},
and this is also the focus of the present work. For a long
biopolymer with zero mean curvature, again it has been
demonstrated that the effect of intrinsic sequence-dependent
curvatures can also be reduced into a simple correction of the
uniform persistence length, either free of external force
\cite{PT07,ZJ08,TTH87,SH95} or under moderate external force
\cite{PN98,BDM98,VATA03}. However, it is argued that sequence
disorder is immaterial for the elasticity of a long DNA under
strong stretching force \cite{MS95,PN98}. In this paper, we prove
it exactly for a two dimensional biopolymer even with a
nonvanishing mean intrinsic curvature. Moreover, it is well known
that the short or intermediate-length DNAs play a more important
role than the long DNAs in biological processes, from DNA
packaging, to transcription, gene regulation and viral packaging
\cite{JW01,RHL95,SLB81}. As a consequence, the effect of intrinsic
sequence-dependent curvatures for short or intermediate-length
biopolymers requires more attention. When the biopolymer is free
of external force and with SRC in intrinsic sequence-dependent
curvatures, it has been shown exactly that such a
three-dimensional (3D) system is equivalent to a system (we will
refer it as the ``equivalent system" henceforth) with a
well-defined (i.e., without randomness) intrinsic mean curvature
and a corrected persistence length \cite{PT07,ZJ08}, irrespective
of its length. In this work, we show that the same conclusion is
also valid in the two-dimensional (2D) case and present the
general solution of the distribution function of the equivalent 2D
system. On the other hand, the effects of intrinsic
sequence-dependent curvatures in a short biopolymer under external
force are not yet known. In this paper, we demonstrate that under
external force, the effect of sequence-disorder for a short
biopolymer is dependent on the average order and the experimental
conditions. Moreover, we find that the results are also dependent
on the boundary conditions (BC), and the short biopolymer looks
softer than what is expected from the ``equivalent system".
Because theoretically a 2D system is relatively easier to study
and experiments on the conformations of biopolymers are often
conducted in a 2D environment \cite{MFFA07,CSM99}, this work will
focus on the 2D system.

The paper is organized as follows. In section II we set up our
model. Section III presents the exact proof of the existence of
the ``equivalent system" and the general distribution function of
the ``equivalent system" for a force-free biopolymer. In section
IV we focus on the conformation and elasticity of the biopolymer
under constant external force. Follows a section discussing the
effects of disorder in a segment dependent curvature in the
constant extension ensemble. Finally, we end the paper with
conclusions and discussions.

\section{The model}
The configuration of a filament with negligible cross section can
be described by the tangent vector, {\bf t}$(s)$, to its contour
line, where $s$ measures the location along the filament. In two
dimensions, {\bf t}$(s)=\{\cos\phi(s), \sin\phi(s)\}$, where the
azimuthal angle $\phi$ is the angle between the $x$-axis and {\bf t}.
The locus of the filament can be found by
\begin{eqnarray}
{\bf r}(s)=\{ x(s),y(s)\}=\int_0^s{\bf t}(u)du.\label{locus1}
\end{eqnarray}
The reduced energy of the filament with intrinsic curvature but
free of external force can be written as:
\begin{eqnarray}
{\cal E}_0[\{\phi(s)\}]\equiv {E\over k_B T}=\int_0^L {k \over
2}[\dot\phi-c(s)]^2ds,\label{energy1}
\end{eqnarray}
where $E$ is the energy, $\dot \phi\equiv d\phi/ds$, $T$ is the
temperature, $k_B$ is the Boltzmann constant, $L$ is the total arc
length of the filament and is a constant in the model so that the
filament is inextensible, $k=l_p/2$ with $l_p$ the 2D bare
persistent length, $c(s)$ is the intrinsic sequence-dependent
curvature. Under a uniaxial applied force $f_x$ (along $x$-axis),
the reduced energy of the filament becomes
\begin{eqnarray}
{\cal E}={\cal E}_0[\{\phi(s)\}]-f\int_0^L \cos\phi
ds,\label{energy2}
\end{eqnarray}
where the reduced force is defined by $f\equiv f_x/k_B T$. When
$c(s)=0$ and $l_p$ is a constant, it returns to the well known WLC
model \cite{KP49,MS94,BMSS94,MS95}. Note that with free boundary
condition, a negative value of $f_{x}$ only extends the polymer in
the negative direction rather than the positive direction. So it does
for a long polymer since the boundary condition becomes
unimportant. Therefore, the sign of the force is meaningless in
these cases. However, it is not the case for a short polymer with
a fixed initial angle.

If both $l_p$ and $c(s)$ are well-defined functions of $s$, a
macroscopic quantity $B_\phi$ is defined as the average with
Boltzmann weights over all possible conformations,
\begin{eqnarray}
B_\phi = {1\over {\cal Z}_k} \int {\cal D}[\phi(s)] B[\{\phi(s)\}]
\text{e}^{-{\cal E}}, \label{mean1}
\end{eqnarray}
where ${\cal Z}_k \equiv \int {\cal D}[\phi]\text{e}^{-{\cal E}}$.
Function $B[\{\phi(s)\}]$ represents different physical
situations. For instance, if $B[\{\phi(s)\}]={\bf t}(s_1)\cdot
{\bf t}(s_2)$, we find the orientational correlation function
(OCF); if $B[\{\phi(s)\}]=|{\bf r}_L-{\bf r}_0|^2$, we obtain the
mean end-to-end distance, where ${\bf r}_L={\bf r}(L)$ and ${\bf
r}_0={\bf r}(0)$; if $B[\{\phi(s)\}]=\delta({\bf R}-\int_0^L{\bf
t}ds)$, we get the distribution function of end-to-end vector; if
$B[\{\phi(s)\}]=\delta({\bf r}_L-{\bf r}_0)\delta[{\bf t}(L)-{\bf
t}(0)]$, we find the looping probability. $B[\{\phi(s)\}]$ is
independent of $k$ and $c(s)$ but can be a very complex function
of $\phi(s)$ and $\dot\phi(s)$.

Eq. (\ref{mean1}) uses $\phi$ as the variable of integration. However, the
variable of integration can be replaced by $\dot{\phi}(s)$, i.e., we
have the following identity (see Appendix 1 for proof)
\begin{eqnarray}
{ \int {\cal D}[\phi(s)] B[\{\phi(s)\}] \text{e}^{-\cal E}\over
\int {\cal D}[\phi(s)] \text{e}^{-\cal E}}={ \int {\cal D}[\dot
\phi(s)] B[\{\phi(s)\}] \text{e}^{-\cal E}\over \int {\cal D}[\dot
\phi(s)] \text{e}^{-\cal E}}. \label{equity1}
\end{eqnarray}

For a biopolymer without correlation on $c(s)$, or with SRC on
$c(s)$ but in the coarse-grained model, the distribution of
$c(s)$, $W(\{c(s)\})$, can be written as a Gaussian distribution
with mean $\bar{c}$, and root-mean squared deviation $1/ \sqrt\alpha$:
\begin{eqnarray}
W(\{c(s)\})= \text{exp}\left[ -\int {\alpha \over 2
}[c(s)-\bar{c}]^2 ds\right]. \label{weight1}
\end{eqnarray}
In this case, we need to average over $c$ for all biopolymers in
the system. Note that averaging can be done in two different
orders, either
\begin{eqnarray}
B\equiv\left< B[\{\phi(s)\}]\right>={1\over {\cal
Z}_\alpha}\int{\cal D}[c(s)]W(\{c(s)\})B_\phi , \label{mean2}
\end{eqnarray}
where ${\cal Z}_\alpha \equiv \int{\cal D}[c(s)]W(\{c(s)\})$, or
\begin{eqnarray}
B'&\equiv&\left< B[\{\phi(s)\}]\right>'\nonumber \\&=&{1\over
{\cal Z}_k} \int {\cal D}[\phi] \left[ {1\over {\cal Z}_\alpha}
\int{\cal D}[c]W(\{c\}) B[\{\phi\}]\text{e}^{-{\cal
E}}\right]\nonumber
\\&=&{1\over {\cal Z}_k} \int {\cal D}[\phi] B[\{\phi\}]\left[
{1\over {\cal Z}_\alpha} \int{\cal D}[c]W(\{c\}) \text{e}^{-{\cal
E}}\right]. \label{mean3}
\end{eqnarray}
Physically, Eq. (\ref{mean2}) corresponds to performing a conformational
or thermal average over an individual sample first, and then a
disorder average over all samples in the system, so we call this a
{\em thermal-first} system. In contrast, in Eq. (\ref{mean3}) a
disorder average over an instantaneous conformation of all samples
is first performed, followed by a conformational average over all
possible conformations, so we refer to the corresponding system as
a {\em disorder-first} system. To find a macroscopic quantity in
the {\em disorder-first} system is equivalent to averaging the
sequence disorder over all samples first to create an
``equivalent" system which is then thermally averaged. So it is
always possible to construct an ``equivalent system" in that case.
However, in experiments and computational simulations {\em
thermal-first} averages are performed. Therefore, this work will
also focus on the {\em thermal-first} system and show that in many
cases the two systems are not equivalent.

\section{Distribution Function for the Force-Free System}
We first present a brief description of the force-free system.
Using the identities Eqs. (\ref{equity1}) and
\begin{eqnarray}
\int{\cal D}[c]W(\{c\}) \text{e}^{-{\cal E}_0}={{\cal Z}_k^0{\cal
Z}_\alpha \over {\cal Z}_{\cal H}^0} \text{e}^{-{\cal
H}_0},\label{eq12}
\end{eqnarray}
and exchanging the order of integration, Eq. (\ref{mean2}) becomes
\begin{eqnarray}
B&=&{1\over {\cal Z}_k^0} \int {\cal D}[\phi]
B[\{\phi(s)\}]{\int{\cal D}[c(s) ] \text{e}^{-\cal E} W(\{ c(s)
\}) \over {\cal Z}_\alpha }
\nonumber \\
&=& {1\over {\cal Z}_{{\cal H}}^0} \int {\cal D}[\phi]
B[\{\phi(s)\}] \text{e}^{-{\cal H}_0}, \label{mean4}
\end{eqnarray}
where
\begin{eqnarray}
{\cal H}_0&=&{1\over 2}\int_0^L
\kappa[\dot{\phi}(s)-\bar{c}]^2ds,\label{energy3}\\
{\cal Z}_k^0 &=& \int {\cal D}[\phi]\text{e}^{-{\cal E}_0}, \text{
}{\cal Z}_{{\cal H}}^0=\int {\cal D}[\phi]\text{e}^{-{\cal H}_0},
\end{eqnarray}
and the effective persistent length,
\begin{eqnarray}
l^{\text{eff}}_p= 2\kappa=2 k \alpha /(k+\alpha). \label{leff}
\end{eqnarray}
Note that Eq. (\ref{mean4}) is valid for any $L$ and even if
$\bar{c}$, $k$ and $\alpha$ are $s$-dependent. Comparing Eqs.
(\ref{mean1}) and (\ref{mean4}), we reach the conclusion that a
system with SRC in $c(s)$ is equivalent to a system with a
well-defined mean intrinsic curvature $\bar{c}$ and a renormalized
persistence length $l_p^{\text{eff}}$. %The randomness in $c(s)$ is
%removed in the equivalent system (with effective energy ${\cal
%H}_0$) so it will be much easier to study.

The same conclusion has been achieved for 3D biopolymers following
similar arguments \cite{ZJ08}, except that in the 2D case we have
to derive Eq. (\ref{equity1}) first due to the convention
of using $\phi$ as integral variable. %This conclusion also agrees
%with what has been found in 3D systems with $k=l_p^s$ and
%$\bar{c}=0$, namely that the randomness of the $c(s)$ can be
%accounted for by replacing $l_p$ with $l_p^{\text{eff}}$ in the
%WLC model, where $1/l_p^{\text{eff}}=1/ l_p+1/l_p^s$
%\cite{PN98,PT07,TTH87,SH95}.
This conclusion also means that the {\em thermal-first} system is
exactly the same as the {\em disorder-first} system in the
force-free case.

From the standard connection between the path integral and the
Schr\"{o}dinger equation, we can find that the partition function
${\cal Z}_{\cal H}^0(\phi(s),s;\phi(s_0),s_0)$ ($\equiv \int {\cal
D}[\phi]\text{e}^{-{\cal H}_0}$) for the system with effective
energy ${\cal H}_0$ satisfies the following partial differential
equation \cite{ZZO00,ZZ07,HK90}
\begin{eqnarray}
{\partial {\cal Z}_{\cal H}^0 \over \partial s}=\left({1\over
2\kappa}{\partial^2 \over \partial \phi^2}-\bar{c}{\partial \over
\partial \phi}\right){\cal Z}_{\cal H}^0. \label{partition1}
\end{eqnarray}
Fixing $\phi(s)$ at $s=s_0$, the boundary condition (BC) becomes
\begin{eqnarray}
{\cal Z}_{\cal
H}^0(\phi,s_0;\varphi,s_0)=\delta[\phi-\varphi],\label{BC3}
\end{eqnarray}
where $\varphi=\phi(s_0)$. In an experiment, one usually takes
$\phi(0)=0$ when $f=0$.

It is straightforward to show that the normalized function (i.e., the
distribution function with $s>s_0$)
\begin{eqnarray}
P(\phi,s;\varphi,s_0)= \sqrt{1\over 2 \pi
A(s,s_0)}\text{e}^{-[\phi-\varphi-\int_{s_0}^s\bar{c}(s)ds]^2/2A(s,s_0)}
\nonumber \\
\label{partition3}
\end{eqnarray}
satisfies Eqs. (\ref{partition1}) and (\ref{BC3}), where
$A(s,s_0)=\int_{s_0}^s ds/\kappa(s)$. Eq. (\ref{partition3}) can
also be derived directly by using a standard path integral
technique \cite{HK90,KLB93} (also see Appendix 2). %The
%orientational correlation function can then be found to be
%($s>s'$)
%\begin{eqnarray}
%&&\left<{\bf t}(s)\cdot {\bf t}(s')\right>_\kappa
%\nonumber \\
%&=&\int_{-\infty}^{\infty}d\phi d\phi'
%P(\phi,s;\phi',s')\cos(\phi-\phi')P(\phi',s';0,0)\nonumber \\
%&=&\text{e}^{-A(s,s')/2}
%\cos\left(\int_{s'}^s\bar{c}(s)ds\right),\label{ocf1}
%\end{eqnarray}
%where $\left< ... \right>_\kappa$ denotes the configurational
%average with effective energy ${\cal H}_0$. When $\kappa$ is a
%constant and $\bar{c}=0$, Eqs. (\ref{partition3}) and (\ref{ocf1})
%lead to the well known result for the WLC model. Therefore,
%Eqs. (\ref{partition3}) is a generalization of the WLC model and
%is valid even when both $\kappa$ and $\bar{c}$ are dependent on $s$.

\section{Conformation and Elasticity of the System Under External Force}

\subsection{On the {\em Disorder-first} System}

When $f\neq 0$, for the {\em disorder-first} system, the derivation leading to
Eq. (\ref{mean4}) can be generalized easily to obtain an equivalent
system with the effective energy
\begin{eqnarray}
{\cal H}={\cal H}_0-f\int_0^L \cos\phi ds, \label{energy4}
\end{eqnarray}
no matter what the force may be since the force term in $\cal E$
is independent of $c(s)$. The equivalent system has been well
studied \cite{PHK05,KMTS07,ZZ07}.

In the three dimensional case, we can also reach a similar
conclusion by a direct generalization of the proof in Ref. \cite{ZJ08}
to find an equivalent system, and an alternative proof for the
three dimensional filament under weak force can be found in Ref.
\cite{PN98}.

\subsection{General Expressions for the {\em Thermal-first} System Under Weak Force}
Note that, mathematically, the conclusion of the existence of an
equivalent system in the force-free case results from the fact that
both ${\cal Z}_k^0$ and ${\cal Z}_\alpha$ are Gaussian integrals
so are independent of $c(s)$ or $\phi$. However, such an argument
fails for the thermal-first system with $f\neq 0$, because ${\cal
Z}_k$ is no longer a Gaussian integral and is dependent on $c$, so
exchanging the order of integration does not simplify the
expression. In other words, there is no simple way to remove the
randomness in $c(s)$ so there is in general no ``equivalent
system"  even under weak force, as we will show exactly below. In
this case, to first order in $f$, as shown in Appendix 3,
\begin{eqnarray}
B&=&{1\over {\cal Z}_\alpha}\int {\cal
D}[\phi]B[\{\phi\}]\text{e}^{F}\left[\int{\cal D}[c]{1\over {\cal
Z}_k}W(\{c\}) \text{e}^{-{\cal E}_0}\right] \nonumber \\
&\approx& B_1-B_2,   \label{B3}\\  \mbox{with} \nonumber \\B_1 &=&
\left< B\right>_\kappa + f\int_0^Lds'\left<
B[\{\phi(s)\}]\cos\phi(s') \right>_\kappa, \label{B11}\\
\mbox{and} \nonumber \\ B_2 &=&
f\int_0^Lds'\text{e}^{-A'(s',0)/2}\left<
B[\{\phi(s)\}] \cos[\gamma(s')+\phi_0]\right>_\kappa, \nonumber \\
\label{B21}
\end{eqnarray}
where
\begin{eqnarray}
A'(s,s_0)&=&\int_{s_0}^s ds/k'(s),\text{ }k'(s)={k (\alpha+k)\over
\alpha+2k},\\
\gamma(s)&=&\int_0^s \dot{\gamma}(s)ds, \text{ }\dot{\gamma}(s)={k
\dot{\phi}+\alpha\bar{c}\over \alpha+k},\label{gamma1}
\end{eqnarray}
and $\phi_0=\phi(0)$ is the initial azimuthal angle. Note that
$\gamma(s)$ is in general dependent on $\phi(s)$  making these
expressions very complex.

We should remind that when $f\neq 0$, $\phi_0$ is not necessarily
zero but is dependent on the experimental conditions. In
experiments, $\phi_0$ may be fixed. In this case, the boundary
condition at $s=0$ is given by Eq. (\ref{BC3}). However,
experiments on stretching biopolymers usually involve attaching
the two ends of the biopolymer to beads, and it does not seem to
be easy to completely prohibit the rotation of the beads. As a
consequence, it may be difficult to fix $\phi_0$. In the extreme
case, the polymer can rotate freely around the origin. This can be
realized by a magnetic tweezer \cite{DC07}. In the more general
case, $\phi_0$ may have a distribution and therefore finally we
need to average over $\phi_0$. It has been reported that different
boundary conditions have considerable effects on the mechanical
response of a homopolymer \cite{ZLJ05,DC07}. In this work, we come
to the same conclusion for a heteropolymer.

On the other hand, for the {\em disorder-first} system, we find
\begin{eqnarray}
B'&=&{1\over {\cal Z}_{\cal H}} \int {\cal D}[\phi] B[\{\phi\}]
\text{e}^{-{\cal H}}
\approx B_1-B_2',\nonumber \\
B_2'&=& f\left< B[\{\phi(s)\}]\right>_\kappa\int_0^Lds\left<
\cos\left(\phi(s)\right)\right>_\kappa
\nonumber \\
&=&f\left< B[\{\phi(s)\}]\right>_\kappa {\cal B}_2',   \label{B2p}\\
{\cal B}_2'&=&\int_0^L ds\text{e}^{-A(s,0)/2}
\cos\left(\phi_0+\int_0^s\bar{c}(s)ds\right),\nonumber
\end{eqnarray}
where ${\cal Z}_{{\cal H}}=\int {\cal D}[\phi]\text{e}^{-{\cal
H}}$.

It is clear that in general $B_2\neq B_2'$ since $\gamma$ is
dependent on $\phi$. It in turn leads to in general $B\neq B'$. In
other words, in general it is impossible to find an equivalent
system for the {\em thermal-first} system under external force.

\subsection{Elasticity of the {\em Thermal-first} System Under Weak Force}

To figure out how serious the effect of the disorder in $c(s)$ or
how large the discrepancy between $B$ and $B'$, we examine the
most interesting and also the simplest case with constant $k$,
$\alpha$ and $\bar{c}=0$. It corresponds to the WLC model and is
often used to describe the entropic elasticity of biopolymers,
such as dsDNA and proteins. Experiments in 3D dsDNA found that
$k\approx 78$nm and $\kappa \approx
45$nm\cite{TTH87,SS90,BFKSDS95,FBSKMSD97}. It follows that
$k\approx 1.7 \kappa $ for DNA. In this case, $\gamma(s)=\kappa
[\phi(s)-\phi_0]/\alpha$, and from Eq. (\ref{B2p}), we can obtain
\begin{eqnarray}
B_2'\approx 2\kappa
f\left(1-\text{e}^{-L/2\kappa}\right)\cos\phi_0\left<
B[\{\phi(s)\}]\right>_\kappa.
\end{eqnarray}

The extension $X$ in the {\em thermal-first} system can be found as
(see Appendix 4)
\begin{eqnarray}
X&\equiv& \left<x\right>-\left<x\right>_{f=0}
\nonumber \\
&\approx&2\kappa f\left[ L-{k(k+3\alpha)\over
k+\alpha}+{k(k+\alpha)\over k-\alpha}
\text{e}^{-L/k}\right.\nonumber
\\&&\left.-{4 \kappa \alpha\over k-\alpha}
\text{e}^{-L/2\kappa}\right]
-{\kappa^2\cos(2\phi_0)\text{e}^{-2L/\kappa}f\over
3(2k+\alpha)(3k+\alpha)}{\cal X},\label{extension1}
\\{\cal X}&=&
6k^2\left( \text{e}^{L/k}-1\right)\nonumber
\\&&+k\alpha\left(9
\text{e}^{2L/\kappa}-16\text{e}^{3L/2\kappa}+12\text{e}^{L/k}-5\right)
\nonumber \\ &&+\alpha^2\left(3
\text{e}^{2L/\kappa}-8\text{e}^{3L/2\kappa}+6\text{e}^{L/k}-1\right).\nonumber
\end{eqnarray}

On the other hand, in the {\em disorder-first} system, the extension $X'$ is:
\begin{eqnarray}
X'&\equiv&\left<x\right>'-\left<x\right>'_{f=0}
\nonumber \\
&\approx&2\kappa f\left[L+\kappa-\kappa
\left(\text{e}^{-L/2\kappa}-2\right)^2\right] +{\kappa^2f\over
3}\cos(2\phi_0)\nonumber \\&&\cdot \left(\text{e}^{-2L/\kappa}
-6\text{e}^{-L/\kappa}+8\text{e}^{-L/2\kappa}-3\right).\label{extension2}
\end{eqnarray}

From Eqs. (\ref{extension1}) and (\ref{extension2}), we can show
that $X\rightarrow X'$ when $\alpha \rightarrow \infty$, as it
should be since in this case the system is free of randomness.
Moreover, for a long polymer, we obtain $X\approx X'\approx 2\kappa
L f$, so that the averaging order is irrelevant for a long
polymer, and agrees with known results
\cite{PN98,BDM98,VATA03}.

However, the discrepancy between the two systems is serious up to
moderate length at finite $\alpha$. From Eqs. (\ref{extension1})
and (\ref{extension2}), we find that the extensions consist
of two terms. The first term is independent of BC or $\phi_0$, but
the second term is dependent on cos$2\phi_0$ so is dependent on
the BC. As a consequence, different boundary conditions have
strong effects on the elasticity up to moderate length. Without loss
in generality, we consider two extreme cases. The first case is to
fix $\phi_0=0$, which gives the most stringent BC effects. The
second is to let $\phi_0$ free so $\left<\cos2\phi_0\right>=0$,
and only the BC-independent term remains. For a better
comparison, we define a ratio of the extensions, $r_e=X/X'$.

\begin{figure}[htbp]
\resizebox{3in}{2.5in}{\includegraphics{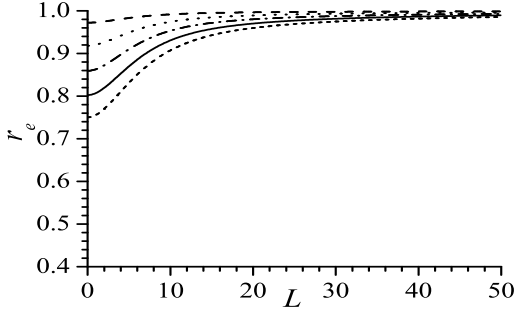}} \caption{The ratio $r_e=X/X'$
vs $L$ for the fixed BC ( $\phi_0=0$). From top to bottom, the
parameters are: $k=1.2\kappa$, $\alpha=6.0\kappa$(dash);
$k=1.4\kappa$, $\alpha=3.5\kappa$ (dot); $k=1.6\kappa$,
$\alpha=2.67\kappa$ (dash dot); $k=1.8\kappa$, $\alpha=2.25\kappa$
(solid); $k=2.0\kappa$, $\alpha=2.0\kappa$ (short dash). The unit
of $L$ is in $\kappa$. }\label{zhouFig2}
\end{figure}

\begin{figure}[htbp]
\resizebox{3in}{2.5in}{\includegraphics{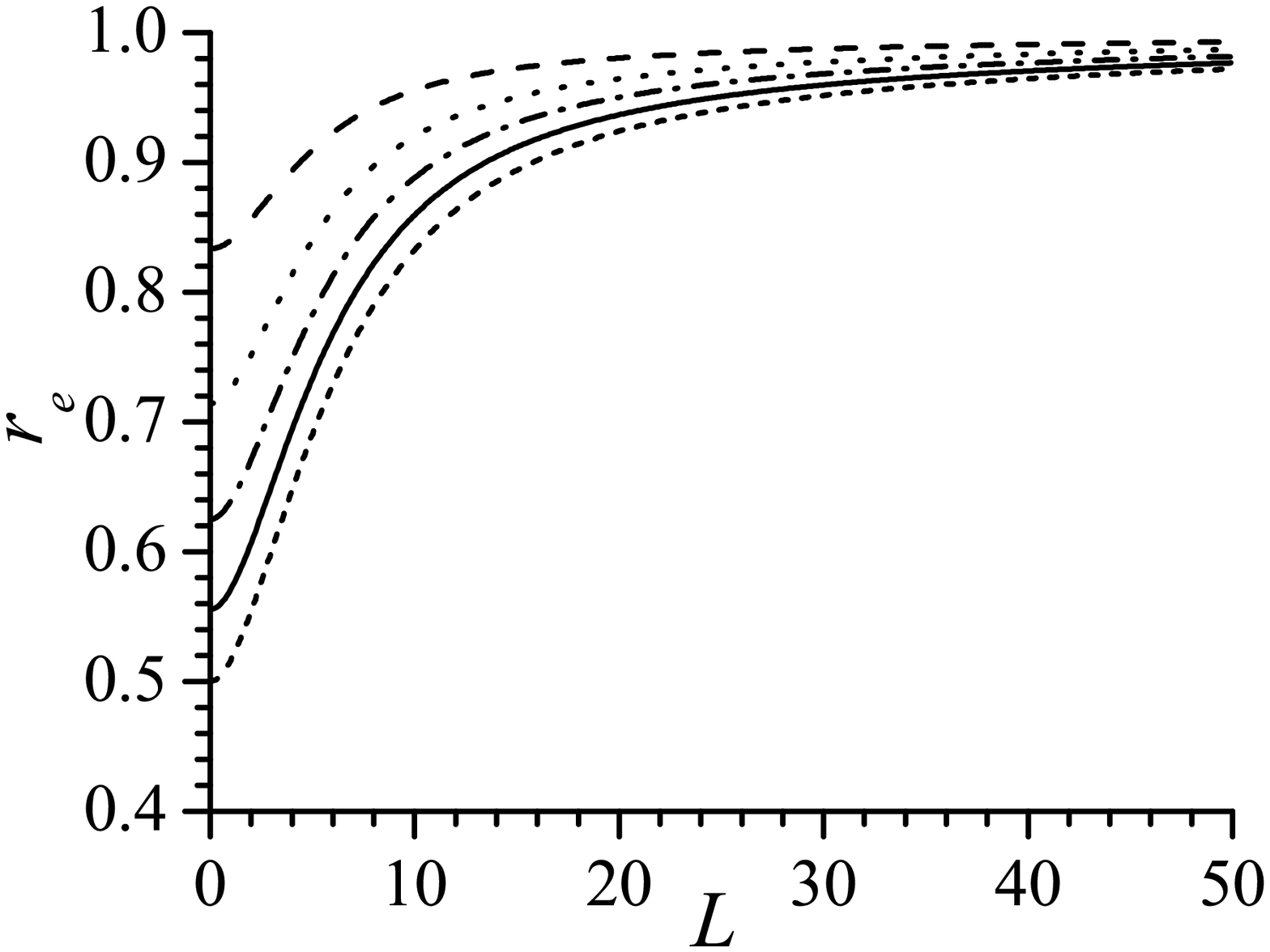}} \caption{The ratio $r_e=X/X'$
vs $L$ for the free BC ($\left<\cos2\phi_0\right>=0$). From top to
bottom, the parameters are: $k=1.2\kappa$, $\alpha=6.0\kappa$
(dash); $k=1.4\kappa$, $\alpha=3.5\kappa$ (dot); $k=1.6\kappa$,
$\alpha=2.67\kappa$ (dash dot); $k=1.8\kappa$, $\alpha=2.25\kappa$
(solid); $k=2.0\kappa$, $\alpha=2.0\kappa$ (short dash). The unit
of $L$ is in $\kappa$.}\label{zhouFig3}
\end{figure}

Figures \ref{zhouFig2} and \ref{zhouFig3} present some typical
results for $r_e$. From these two figures, we can see that $r_e$
increases monotonously with increasing $L$, and in general the
disorder in $c(s)$ makes the biopolymer in the {\em thermal-first}
system softer, i.e. $r_e<1$, than the ``equivalent"
system, and the effects may be still rather serious up to the
length of about 20 $\kappa$. Furthermore, we also find that the
results are sensitive to BC and randomness of $c(s)$. At first,
the effect is much more serious (maybe about twice) in the free
BC (Fig. \ref{zhouFig3}) than in fixing BC (Fig. \ref{zhouFig2}).
Second, the effect is getting stronger with increasing randomness,
i.e. with decreasing $\alpha$. Especially, at $k\approx 1.7\kappa$
which corresponds to the experimental value of DNA, we find that
$X$ can be only about half of $X'$ for $L\sim \kappa$, and 20\%
smaller than $X'$ up to $L\approx 10 \kappa $ or about 1500 bp.
That means DNA is more likely to be in coil state or appears
softer and has a smaller apparent persistent length than that in
the ``equivalent" system up to a rather long length. Finally,
these results are independent of the external force, so that a
small force may produce a considerable effect. This fact may be
important in a stretching experiment for a short polymer since it
suggests that the interaction between experimental apparatus and
polymer may affect the results seriously.

\subsection{Orientational Correlation Function (OCF) and End-to-end
Distance for the {\em Thermal-first} System Under Weak Force}

Let $B[\{\phi\}]={\bf t}(s)\cdot {\bf
t}(s')=\cos[\phi(s)-\phi(s')]$, from Eqs.
(\ref{B3})-(\ref{gamma1}), we obtain the orientational correlation
function ($s>s'$) for the {\em thermal-first} system and under
weak external force (see Appendix 5)

\begin{eqnarray}
&&\left<{\bf t}(s)\cdot {\bf t}(s')\right> \approx
\text{e}^{-(s-s')/2\kappa}+{\kappa \cos\phi_0 f  \over
3(3k+\alpha)(k-\alpha)} {\cal S},\nonumber \\ {\cal S}&=&-6k
(k-\alpha)\text{e}^{-2[(\alpha+2k)s+(\alpha +3k)s']/2k\alpha}
\nonumber \\
&&+8\alpha(2k+\alpha)\text{e}^{-s/2\kappa} -6k
(3k+\alpha)\text{e}^{-[2\alpha
s+(k-\alpha)s']/2k\alpha}\nonumber \\
&&+(k-\alpha) (3k+\alpha)\left[3\text{e}^{-[(\alpha+k)L+(\alpha
+3k)(s-s')]/2k\alpha} \right.\nonumber \\
&&-3\text{e}^{-[L+3(s-s')]/2\kappa}+6\text{e}^{-s'/2\kappa}
-3\text{e}^{-(L+s'-s)/2\kappa} \nonumber \\
&& \left.
+2\text{e}^{-(4s-3s')/2\kappa}+3\text{e}^{-[(\alpha+k)L+(\alpha
-k)(s-s')]/2k\alpha}\right].\label{ocf2}
\end{eqnarray}

In contrast, in the {\em disorder-first} system,
\begin{eqnarray}
&&\left<{\bf t}(s)\cdot {\bf t}(s')\right>' \approx
\text{e}^{-(s-s')/2\kappa} +{1\over 3}\kappa f\cos\phi_0{\cal
S}',\nonumber
\\{\cal
S}'&=&6\text{e}^{-s'/2\kappa}-8\text{e}^{-s/2\kappa}
+6\text{e}^{-(L+s-s')/2\kappa}+2\text{e}^{(3s'-4s)/2\kappa}
\nonumber
\\&& -3\text{e}^{-[L+3(s-s')]/2\kappa}-3
\text{e}^{-(L+s'-s)/2\kappa}. \label{ocfp}
\end{eqnarray}

The end-to-end distance can be found as
\begin{eqnarray}
R^2 &\approx&4\kappa L \left[ 1-{2\kappa \over
L}\left(1-\text{e}^{-L/2\kappa}\right)\right]\nonumber
\\ &&+{4\kappa^2
\cos\phi_0 f \over
9(k-\alpha)^2(k+\alpha)(2k+\alpha)(3k+\alpha)^2}{\cal
Y},\label{endToEnd1} \\ \mbox{with} \nonumber \\
{\cal Y}&=&18(k-\alpha)^2(k+\alpha)(2k+\alpha)(3k+\alpha)^2L
\nonumber
\\ &&-9k(k-\alpha)^2(3k+\alpha)^2(4k^2+11\alpha^2+22k\alpha)\nonumber
\\ &&-16\alpha (2k+\alpha )^2[3(k-\alpha)(k+\alpha)(3k+\alpha)L
\nonumber
\\ &&+4k\alpha (5k+\alpha)(3k-\alpha)]\text{e}^{-L/2\kappa}\nonumber
\\ &&+18k \alpha
(k+\alpha)^3(k-\alpha)^2\text{e}^{-(1/k+2/\alpha)L}\nonumber
\\ && +18k
(k+\alpha)^3(2k+\alpha)(3k+\alpha)^2\text{e}^{-L/k}\nonumber
\\ &&-k\alpha
(k-\alpha)^2(2k+\alpha)(3k+\alpha)^2\text{e}^{-2L/\kappa}.\nonumber
\end{eqnarray}

On the other hand
\begin{eqnarray}
R'^2&\approx&4\kappa L \left[ 1-{2\kappa \over
L}\left(1-\text{e}^{-L/ 2\kappa}\right)\right] +{4\over 9}\kappa^2
\cos\phi_0 f {\cal Y}'\label{endToEnd2}\nonumber \\
\\ \mbox{with} \ \ {\cal Y}'&=&
18L-99\kappa+16(3L+4\kappa)\text{e}^{-L/2\kappa}\nonumber \\ &&
+36\kappa\text{e}^{-L/\kappa}
-\kappa\text{e}^{-2L/\kappa}.\nonumber
\end{eqnarray}

When $\alpha \rightarrow \infty$, i.e. a system without
randomness, we obtain $R^2=R'^2$ as it should be. Moreover, $L
\rightarrow \infty$ also leads to $R^2\rightarrow R'^2$, again
supporting the conclusion that the order in averaging is irrelevant
for a long polymer \cite{PN98,BDM98,VATA03}.

\begin{figure}[htbp]
\resizebox{3in}{2.5in}{\includegraphics{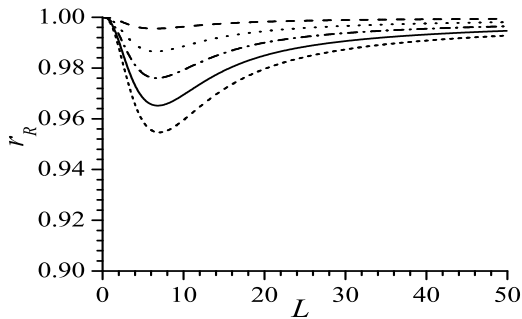}} \caption{The ratio $r_R=R^2/R'^2$
vs $L$. $f=0.5/\kappa$ and $\phi_0=0$ (the fixed BC) for all
curves. From bottom to top, the parameters are: $k=1.2\kappa$,
$\alpha=6.0\kappa$ (dash); $k=1.4\kappa$, $\alpha=3.5\kappa$
(dot); $k=1.6\kappa$, $\alpha=2.67\kappa$ (dash dot);
$k=1.8\kappa$, $\alpha=2.25\kappa$ (solid); $k=2.0\kappa$,
$\alpha=2.0\kappa$ (short dash). The unit of $L$ is in
$\kappa$.}\label{zhouFig4}
\end{figure}

Moreover, from Eqs. (\ref{endToEnd1}) and (\ref{endToEnd2}), we
find that the end-to-end distance can also be divided into two
terms. The first term is the force free term and is independent of
BC, but the second term is dependent on both BC and force.
Especially, for free BC ($\left<\cos2\phi_0\right>=0$), the force
has not effect at all, this is quite different from the
force-extension relation. Furthermore, with finite
$\left<\cos2\phi_0\right>$, $R^2$ is also smaller than $R'^2$,
similar to the force-extension relation. However, the ratio
$r_R=R^2/R'^2$ is no longer a monotonic function of $L$, but has a
minimum at $L\approx 6\kappa$. Moreover, the discrepancy between
$R$ and $R'$ increases with increasing $f$ (see Eqs. (\ref{endToEnd1})
and (\ref{endToEnd2})). Note that our results
are valid only at weak force, and for DNA it means that we require
$k_BTf<f_c\equiv k_BT/\kappa $. Taking the generally accepted
value $\kappa\approx$ 50 nm, we have $f_c\approx 0.08pN$. Fig.
(\ref{zhouFig4}) shows some typical results at $f=0.5/\kappa$, which
corresponds to an external force of about 0.04 pN. From Fig.
(\ref{zhouFig4}), we can see that the discrepancy between $R^2$
and $R'^2$ is much smaller than that for extensions. This is
because the force free term dominates the value of $r_R$. The fact
that the disorder in $c(s)$ has quite different effects on the
extension and the end-to-end distance may be important in
experiments.

\subsection{Elasticity of the Long Biopolymer Under Large Force}
Again we assume constant $k$, $\alpha $ in this part, but allow a
nonvanishing $\bar{c}(s)$. In the large force limit the filament is
nearly straight, thus {\bf t}$(s)$ is nearly pointing along the
direction of the force. That means $\phi(s)\approx 0$ and the
reduced energy becomes
\begin{eqnarray}
{\cal E}\approx \int_0^L {k \over 2}[\dot\phi-c(s)]^2ds+f \int_0^L
{1 \over 2}\phi^2 ds,   \label{energy11}
\end{eqnarray}
where we have dropped a constant term $-fL$. For a very long
filament, we can use periodic boundary conditions with negligible
error, so take $q_n=2\pi n/L$ with integer $n$, and expand $\phi$,
$c(s)$ and $\bar{c}(s)$ as Fourier series
\begin{eqnarray}
\phi(s)&=&\sum_{n=1}^{\infty}a_n\sin(q_ns),
\text{ }\dot{\phi}(s)=\sum_{n=1}^{\infty}a_nq_n\cos(q_ns),
\nonumber \\ \\
c(s)&=&\sum_{n=0}^{\infty}c_n\cos(q_ns),\text{ }
\bar{c}(s)=\sum_{n=0}^{\infty}\bar{c}_n\cos(q_ns).
\end{eqnarray}
Note that to use a sine series for $\phi(s)$ is reasonable since
$\phi_0=\phi(L)=0$. But in general $c(L)\neq 0$ so we cannot use
sine series for it. We can also use the full Fourier series for
$c(s)$ and $\bar{c}(s)$, but it is straightforward to show that
the sine part in the full Fourier series make no contribution at
all so we disregard it. Using the orthogonality property of
Fourier modes, we can reexpress the energy and extension as
\begin{eqnarray}
{\cal E}&\approx& {kc_0^2L\over 2}+{kL\over
4}\sum_{n=1}^{\infty}(a_n q_n-c_n)^2+{fL\over
4}\sum_{n=1}^{\infty}a_n^2,\nonumber
\\
&=&{kc_0^2L\over 2}+{L\over
4}\sum_{n=1}^{\infty}\left[(kq_n^2+f)d_n^2+{kf \over
kq_n^2+f}c_n^2\right], \nonumber \\ \label{energy12}\\ \text{with} \nonumber
\\
d_n&=&a_n-{kq_nc_n\over kq_n^2+f},\\ \text{and} \nonumber
\\x&=&\int_0^L\cos\phi ds
\approx L\left(1-{1\over 4}\sum_{n=1}^{\infty}a_n^2\right)
\nonumber
\\&=&x_1+x_2-{L\over
2}\sum_{n=1}^{\infty}{kq_nc_n\over
kq_n^2+f}d_n,\label{extension12}\\\text{where} \nonumber
\\
x_1&=&L\left[1-{1\over 4}\sum_{n=1}^{\infty}d_n^2\right], \\
x_2&=&-{L\over 4}\sum_{n=1}^{\infty}{(kq_n)^2\over
(kq_n^2+f)^2}c_n^2\label{extension13}.
\end{eqnarray}
From Eqs. (\ref{mean1}) and (\ref{energy12})-(\ref{extension13}),
we see that in the {\em thermal-first} system
$\left<x\right>=\left<x_1\right>+\left<x_2\right>$. $x_1$ is
independent of $c(s)$, and we recover the well known result for
the WLC model \cite{PHK05,KMTS05,ZZ07},
\begin{eqnarray}
{\left<x_1\right>\over L}=1-{1\over 2\sqrt{fl_p}}\label{x1}.
\end{eqnarray}
In contrast, $x_2$ is independent of $\phi(s)$ and is determined
by $c(s)$. Rewriting $W(\{c(s)\})$ as
\begin{eqnarray}
W(\{c(s)\})=\text{e}^{-{\alpha L\over
2}\left[(c_0-\bar{c}_0)^2+{1\over
2}\sum_{n=1}^{\infty}(c_n-\bar{c}_n)^2\right]},
\end{eqnarray}
we obtain
\begin{eqnarray}
\left<x_2\right>&=&-{L\over 4}\sum_{n=1}^{\infty}{(kq_n)^2\over
(kq_n^2+f)^2}\left<c_n^2\right>\nonumber \\
&=&-{L\over 4}\sum_{n=1}^{\infty}{(kq_n)^2\over
(kq_n^2+f)^2}\left(\bar{c}_n^2+{2\over \alpha L}\right)\nonumber \\
&\approx &-{L\over 4}\sum_{n=1}^{\infty}{(kq_n)^2\over
(kq_n^2+f)^2}\bar{c}_n^2.\label{x2}
\end{eqnarray}

On the other hand, replacing $k$ and $c_n$ by $\kappa$ and
$\bar{c}_n$ respectively in Eqs.
(\ref{energy12})-(\ref{extension13}), we find that in the
{\em disorder-first} system the extension becomes
\begin{eqnarray}
{\left<x\right>'\over L}={\left<x_1\right>'+x_2\over L}=1-{1\over
2\sqrt{f\l_p^{\text{eff}}}}+{x_2\over L}\label{x1p}.
\end{eqnarray}
From Eqs. ({\ref{extension12})-({\ref{x1p}), we find that
replacing $k$ by $\kappa$ one goes from the {\em thermal-first} system
to the {\em disorder-first} system. It is also interesting to note that
$\alpha$ or the width of the distribution of $c(s)$ plays no role
in the extension in both systems. Furthermore, we find that when
$\bar{c}$ is a constant, vanishing or nonvanishing,
$\left<x_2\right>\approx 0$ since all $\bar{c}_n=0$ if $n\ge 1$.
On the hand hand, considering the special case with $\bar{c}=\sigma
$e$^{-\lambda s}$, in the {\em thermal-first} system we have
\begin{eqnarray}
{\left<x_2\right>\over L} &\approx &-{1\over
4}\sum_{n=1}^{\infty}{(kq_n)^2\over
(kq_n^2+f)^2}\bar{c}_n^2\nonumber \\&=& -{\lambda \sigma \over
2L}\sum_{n=1}^{\infty}{(kq_n)^2\over
(kq_n^2+f)^2(q_n^2+\lambda^2)}\nonumber\\
&\approx& -{\lambda \sigma k^2\over
4\pi}\int_0^{\infty}{q^2dq\over
(kq^2+f)^2(q^2+\lambda^2)}\nonumber \\&=& -{\lambda \sigma
k^2\over 16\sqrt{f
k}\left(\sqrt{f}+\lambda\sqrt{k}\right)^2}.\label{x22}
\end{eqnarray}

A long biopolymer in general has a vanishing or small mean
curvature, or in other words in general $\sigma $ and $\lambda$
are small, so that $x_2$ is negligible in either {\em
thermal-first} or {\em disorder-first} systems. Since the
corrections of $\bar{c}$ are negligible in both constant and fast
decay cases, we can conclude safely that $x_2$ is always
negligible. In the {\em thermal-first} system, it means that
sequence-disorder has no effect, which agrees with Marko and
Siggia's argument that under a strong stretching force, disorder
in sequence is immaterial for elasticity \cite{MS95,PN98}. Note
that the effect of the sequence-disorder has been absorbed into
$\kappa$ in the disorder-first system, the result in this part
provides another evidence of the non-equivalence of the two
systems.

\section{On the Constant-extension Ensemble}

Up to now, our discussions are based on a constant external force,
or in the constant-force ensemble. However, the experiments may be
performed with fixed ends, or in the constant-extension ensemble. For
a long polymer, it is believed that these two ensembles should
yield the same mechanical properties. However, it has been
reported that the two ensembles are not always equivalent for
a short polymer. It is therefore interesting to ask whether sequence-disorder
has the same effects in the two ensembles. In the
constant-extension ensemble, without sequence disorder, the
partition function is
\begin{eqnarray}
{\cal Z}_e&=&\int{\cal D}[\phi] \delta({\bf r}_L-{\bf
b})\text{e}^{-{\cal E}_0} \nonumber \\ &=&\int {\cal
D}[\phi]\delta\left(\int_0^L {\bf t} ds-{\bf
b}\right)\text{e}^{-{\cal E}_0}, \label{ze1}
\end{eqnarray}
where $\bf b$ is the constant end-to-end vector. With sequence
disorder, the {\em disorder-first} average for the ensemble
becomes
\begin{eqnarray}
B'&=&{1 \over {\cal Z}_e}\int{\cal D}[\phi]\left[{1 \over {\cal
Z}_\alpha}
 \int{\cal D}[c]W(\{c\})\delta({\bf r}_L-{\bf b})\text{e}^{-{\cal E}_0}B[\{\phi\}]
 \right]\nonumber \\
&=&{1 \over {\cal Z}_e}\int{\cal D}[\phi]\delta({\bf r}_L-{\bf
b})B[\{\phi\}]\left[{1 \over {\cal Z}_\alpha}
 \int{\cal D}[c]W(\{c\})\text{e}^{-{\cal E}_0}\right]\nonumber \\
&=&{{\cal Z}_k^0 \over {\cal Z}_e {\cal Z}_{\cal H}^0} \int{\cal
D}[\phi]\delta({\bf r}_L-{\bf b})B[\{\phi\}]\text{e}^{-{\cal
H}_0},
\end{eqnarray}
where we have used Eq. (\ref{eq12}) again. Now considering the
special case $B[\{\phi\}]=1$, which results in $B'=1$, and
\begin{eqnarray}
{{\cal Z}_k^0 \over {\cal Z}_e {\cal Z}_{\cal H}^0}={1\over
\int{\cal D}[\phi] \delta({\bf r}_L-{\bf b})\text{e}^{-{\cal
H}_0}}, \label{lambda1}
\end{eqnarray}
leads us to the result,
\begin{eqnarray}
B' &=&{\int{\cal D}[\phi]\delta({\bf r}_L-{\bf
b})B[\{\phi\}]\text{e}^{-{\cal H}_0}\over \int{\cal D}[\phi]
\delta({\bf r}_L-{\bf b})\text{e}^{-{\cal H}_0}} , \label{meanE1}
\end{eqnarray}
Therefore, we reach the conclusion that in the constant-extension
ensemble for the {\em disorder-first} system we can still find an
``equivalent system" with the effective energy ${\cal H}_0$.

On the other hand, the {\em thermal-first} average for the
constant-extension ensemble can be written
\begin{eqnarray}
B&=&{1 \over {\cal Z}_\alpha}\int{\cal D}[c]W(\{c\})B_e
,\\
B_e&=&{1 \over {\cal Z}_e}\int {\cal
D}[\phi]B[\{\phi\}]\delta({\bf r}_L-{\bf b})\text{e}^{-{\cal
E}_0}.\label{meanE2}
\end{eqnarray}
From Eq. (\ref{ze1}), we find that due to the existence of the
$\delta({\bf r}_L-{\bf b})$ term, ${\cal Z}_e$ is in general
dependent on $c(s)$, so that an exchange in the order of integration
cannot simplify the expression for $B$. In other words, the
existence of an ``equivalent system" for the {\em thermal-first} system
in the constant-extension ensemble is still an open question. From our
experience in the constant-force ensemble, such an ``equivalent
system" does not exist.

Our discussions in this section can be directly generalized to
the three dimensional system, so it completes and reassess the results of our
previous work \cite{ZJ08}.

\section{Conclusions and discussions}

In summary, we present a rigorous proof that when free of external
force, a 2D semiflexible biopolymer without correlation or with
SRC in intrinsic curvatures $c(s)$ is equivalent to a system with
a well-defined intrinsic curvature and a renormalized persistence
length. We obtain exact expressions for the distribution function
of the equivalent system. These conclusions can simplify
theoretical studies of semiflexible biopolymers, since the
disorder in $c(s)$ is completely removed in the equivalent system.
For the system under external force, we find that the effect of
sequence-disorder is dependent on the order in which the averaging
is done or the experimental conditions. In the {\em
disorder-first} system, it is always possible to find an
``equivalent system", no matter the external force, the length of
the polymer, the statistical ensemble or the dimension of the
system. However, in the {\em thermal-first} system, there is in
general no ``equivalent system" for a biopolymer up to moderate
length. Physically, this is because in the thermal-first system
the intrinsic curvatures favor defects such as kinks, buckles and
loops. To straighten these defects costs extra energy and
therefore requires a larger force. In contrast, the disorder-first
system erases these extra defects before the application of the
force. We find the closed-form expression for the force-extension
relationship for the elasticity of a long biopolymer under a
strong stretching force. In the {\em thermal-first} system, we
show exactly that in this case sequence disorder is immaterial
even if the biopolymer has a nonvanishing mean intrinsic
curvature. Moreover, we find that in the {\em thermal-first}
system, the results are also dependent on the boundary conditions,
and the sequence-dependent effects are much more serious in the
case of free BCs than for fixed BCs. Meanwhile, the results are
dependent on the degree of randomness and the larger the
randomness, the more serious the effect. Our results suggest that
the short biopolymer may be much softer so has a smaller apparent
persistent length than what the ``equivalent system" in a
mechanical experiment would predict. This fact implies that in
experiments the interaction between experimental apparatus and
polymer, though may be weak, may affect the results seriously for
a short polymer. Furthermore, our results suggest that the effects
of sequence-disorder is dependent upon the quantity measured and
how it is measured. We should note that due to the existence of an
``equivalent system", the {\em disorder-first} system is much
simpler in theoretical studies. However, it is difficult to
realize the {\em disorder-first} system in experiment.

On the other hand, we considered weak forces and large forces
but not intermediate size forces in this work,
but we can expect that in this case there will
also not be an ``equivalent system" even for a long polymer since this is
the case for the system under a large force. How the elasticity of a
system goes from an ``equivalent system" when free of force to a
disorder free system under a large force would be an important
question to address. We also do not consider the system with LRC in $c(s)$,
which deserves further investigation.

It should also be noted that this work focussed on 2D systems.
However, whether the results apply to 3D systems is an intriguing
question. Mathematically we should reach similar conclusions since
in the thermal-first system exchanging the order of integration
does not simplify the problem. However, physically there exist
some fundamental distinctions between the 3D case and the 2D case.
At first, the much stronger fluctuations in the 3D system may
dominate the intrinsic disorder so the effect may be suppressed.
This would explain why the disorder in $c(s)$ has a much smaller
effect on the end-to-end distance than on the extension. Moreover,
the geometry of a polymer with natural curvature is also very
different in the 2D and 3D systems. For example, the looped
configuration in the 2D system cannot undergo an out-of-plane
buckling that would eliminate loops, but a 3D polymer will exhibit
this behavior. Furthermore, in the 2D case it is much easier to
form large defects which would be responsible for a larger
decrease in extension. Therefore, the difference in the
thermal-first or disorder-first ordering may be reduced in a 3D
system. But finally let us point out that the studies of the
conformations of biopolymers are often performed in a 2D
environment (e.g., see \cite{MFFA07,CSM99}), so our main findings
should be instructive.

\section*{Appendix 1: Proof of Eq. (\ref{equity1})}

Using the standard path integral methods \cite{KLB93}, for
arbitrary function ${\cal F}[\{\phi(s)\}]$, we can write
\begin{eqnarray}
\int {\cal D}[\phi(s)] {\cal F}[\{\phi(s)\}] &\propto& \lim_{N
\rightarrow \infty } \prod_{j=1}^{N-1}\int d\phi_j{\cal
F}[\{\phi_j\}],\label{part6}\\
\int {\cal D}[\dot \phi(s)] {\cal F}[\{\phi(s)\}] &\propto&
\lim_{N \rightarrow \infty } \prod_{j=1}^{N-1}\int {1\over
\epsilon}d\xi_j{\cal F}[\{\phi_j\}] ,\label{part7}
\end{eqnarray}
where $\epsilon=L/N$, $\phi_j=\phi[(j-1)\epsilon ]$ is the
discretized $\phi(s)$, $\xi_j=\phi_j-\phi_{j-1}$, and $\dot\phi$
in $B$ and $\cal E$ must be replaced by
$(\phi_{j+1}-\phi_j)/\epsilon$. The Jacobian determinant,
$J=|\partial \xi /\partial \phi |$, of $\xi '$s with respect to
$\phi '$s is a constant, therefore,
\begin{eqnarray}
\int {\cal D}[\dot \phi(s)] {\cal F}[\{\phi(s)\}] \propto \lim_{N
\rightarrow \infty }{J\over \epsilon^{N-1}} \prod_{j=1}^{N-1}\int
d\phi_j{\cal F}[\{\phi_j\}].\label{part8}
\end{eqnarray}
Now taking ${\cal F}=B[\{\phi(s)\}]$e$^{-\cal E}$ or e$^{-\cal
E}$, from Eq. (\ref{part8}) we obtain Eq. (\ref{equity1}),
\begin{eqnarray}
{ \int {\cal D}[\phi(s)] B[\{\phi(s)\}] \text{e}^{-\cal E}\over
\int {\cal D}[\phi(s)] \text{e}^{-\cal E}}={ \int {\cal D}[\dot
\phi(s)] B[\{\phi(s)\}] \text{e}^{-\cal E}\over \int {\cal D}[\dot
\phi(s)] \text{e}^{-\cal E}}.
\end{eqnarray}
Intuitively, both sides in the above equation are averages over
all possible configurations so they are expected to be equivalent.

\section*{Appendix 2: A direct derivation of Eq. (\ref{partition3})}

In this appendix, we will use the standard path integral method
\cite{KLB93} to derive Eq. (\ref{partition3}) since it is useful.
To account the more general case of $s_0\neq 0$, we rewrite ${\cal
H}_0$ as
\begin{eqnarray}
{\cal H}_0&=&{1\over 2}\int_{s_0}^L
\kappa[\dot{\phi}(s)-\bar{c}(s)]^2ds.
\end{eqnarray}
For large $N$, the path integral can be approximated as
\begin{widetext}
\begin{eqnarray}
{\cal Z}_{\cal H}^0 \approx C\prod_{j=1}^{N-1}\int
d\phi_j\text{exp}\left\{-{\epsilon\over 2}\sum_{j=0}^{N-1}\kappa_j
\left[{\phi_{j+1}-\phi_j\over \epsilon}-\bar{c}_j\right]^2
\right\}=C\prod_{j=1}^{N-1}\int d\phi_j\text{exp}\left\{-{1\over
2\epsilon}\sum_{j=0}^{N-1}\kappa_j
\left[\phi_{j+1}-\phi_j-\bar{c}_j\epsilon\right]^2 \right\},
\end{eqnarray}
where $C$ is a constant, $\epsilon=(L-s_0)/N$,
$\phi_j=\phi[s_0+(j-1)\epsilon ]$,
$\kappa_j=\kappa[s_0+(j-1)\epsilon ]$ and
$\bar{c}_j=\bar{c}[s_0+(j-1)\epsilon ]$ are discretized $\phi(s)$,
$\kappa(s)$ and $\bar{c}(s)$, respectively, and $\dot\phi$ in
${\cal H}_0$ is replaced by $(\phi_{j+1}-\phi_j)/\epsilon$. Now
using the identity
\begin{eqnarray}
\int_{-\infty}^\infty dx\text{e}^{-a(x-x_1)^2-b(x_2-x)^2}%\nonumber \\&=&
=\sqrt{\pi \over a+b}\text{exp}\left[-{1 \over
1/a+1/b}(x_1-x_2)^2\right],
\end{eqnarray}
we obtain
\begin{eqnarray}
{\cal Z}_{\cal H}^0 &\approx& C\sqrt{2\pi \epsilon\over
\kappa_0+\kappa_1}\int d\phi_2d\phi_3\cdot \cdot \cdot
d\phi_{N-1}\text{exp}\left\{-{1\over
2\epsilon}\sum_{j=2}^{N-1}\kappa_j
\left[\phi_{j+1}-\phi_j-\bar{c}_j\epsilon\right]^2
\right\}\nonumber
\\&&\cdot
\text{exp}\left\{-{1 \over 2\epsilon/ \kappa_0+2\epsilon/
\kappa_1}[\phi_2-\phi_0-(\bar{c}_0+\bar{c}_1)\epsilon]^2\right\}\nonumber
\\&=&C\sqrt{2\pi \epsilon\over
\kappa_0+\kappa_1}\sqrt{2\pi \epsilon\over \kappa'_1+\kappa_2}\int
d\phi_3d\phi_4\cdot \cdot \cdot
d\phi_{N-1}\text{exp}\left\{-{1\over
2\epsilon}\sum_{j=3}^{N-1}\kappa_j
\left[\phi_{j+1}-\phi_j-\bar{c}_j\epsilon\right]^2 \right\}
\nonumber
\\&&\cdot\text{exp}\left\{-{1 \over
2\epsilon/ \kappa'_1+2\epsilon/
\kappa_2}[\phi_3-\phi_0-(\bar{c}_0+\bar{c}_1+\bar{c}_2)\epsilon]^2\right\}\nonumber
\\&=&...={\cal C}\text{ exp}\left\{-{1 \over
2\sum_{j=0}^{N-1}\epsilon/
\kappa_j}\left[\phi_N-\phi_0-\sum_{j=0}^{N-1}\bar{c}_j\epsilon\right]^2\right\},
\end{eqnarray}
where $1/\kappa'_1\equiv 1/\kappa_0+1/\kappa_1$ and ${\cal C}$ is
a new constant. Now let $N\rightarrow \infty$, we obtain
\begin{eqnarray}
{\cal Z}_{\cal H}^0={\cal C}\text{ exp}\left\{-{1 \over
2A(L,s_0)}\left[\phi(L)-\phi(s_0)-\int_{s_0}^Lds\bar{c}(s)\right]^2\right\}.
\label{a1}
\end{eqnarray}
Normalizing the above equation we obtain ${\cal C}=1/\sqrt{2\pi
A(L,s_0)}$, and recover Eq. (\ref{partition3}).

\section*{Appendix 3: Derivation of Eqs. (\ref{B3})-(\ref{gamma1})}
When force is small, expanding e$^{-\cal E}$ about $f=0$, we
obtain
\begin{eqnarray}
{\cal Z}_k &=& \int {\cal D}[\phi]\text{e}^{-\cal E}\approx
{\cal Z}_k^0 (1+f Q),\\
B&=&{1\over {\cal Z}_\alpha}\int {\cal
D}[\phi]B[\{\phi\}]\text{e}^{F}\left[\int{\cal D}[c]{1\over {\cal
Z}_k}W(\{c\}) \text{e}^{-{\cal E}_0}\right] %\nonumber \\ &
\approx {1\over {\cal Z}_\alpha {\cal Z}_k^0}\int {\cal
D}[\phi]B[\{\phi\}]\text{e}^{F}\left[\int{\cal D}[c]W(\{c\})
(1-fQ)\text{e}^{-{\cal E}_0}\right] \nonumber \\
&=& B_1-B_2, \label{mean5}
\end{eqnarray}
where
\begin{eqnarray}
Q&=&\int_0^Lds\left< \cos\phi(s) \right>_k^0,
\text{ }F\equiv f\int_0^L \cos\phi ds,\\
B_1&\equiv& {1\over {\cal Z}_\alpha {\cal Z}_k^0}\int {\cal
D}[\phi]B[\{\phi\}]\text{e}^{F}\left[\int{\cal D}[c]W(\{c\})
\text{e}^{-{\cal E}_0} \right] %\nonumber \\ &=&
={1\over {\cal Z}_{{\cal H}}^0}\int {\cal
D}[\phi]B[\{\phi\}]\text{e}^{-\cal H}\nonumber \\
&\approx& \left< B\right>_\kappa +
f\int_0^Lds'\left< B[\{\phi(s)\}]\cos\phi(s') \right>_\kappa, \label{B11A}\\
B_2 &=&{f\over {\cal Z}_\alpha {\cal Z}_k^0}\int {\cal
D}[\phi]B[\{\phi\}]\text{e}^{F}R %\nonumber \\ &
\approx {f\over {\cal Z}_\alpha {\cal Z}_k^0}\int {\cal
D}[\phi]B[\{\phi\}]{\cal R},\label{B21A}\\
{\cal R}&=&\int{\cal D}[c]W(\{c\}) Q\text{e}^{-{\cal
E}_0},\label{mean6}
\end{eqnarray}
and $\left< ... \right>_k^0$ denotes the ensemble average with
energy ${\cal E}_0$
\begin{eqnarray}
\left< ... \right>_k^0&\equiv& {1\over {\cal Z}_k^0} \int{\cal
D}[\phi(s)](...) \text{e}^{-{\cal E}_0}.
\end{eqnarray}
Furthermore,
\begin{eqnarray}
{\cal R}&=&\int_0^Lds'\left[ \int{\cal D}[c]W(\{c\})
\text{e}^{-{\cal E}_0[\{\phi(s)\}]}{1\over {\cal Z}_k^0} \int
{\cal D}[\phi']\cos[\phi'(s')]\text{e}^{-{\cal
E}_0[\{\phi'(s)\}]}\right]
\nonumber \\
&=&\int_0^Lds'{1\over {\cal Z}_k^0}\left[ \int {\cal
D}[\phi']\cos[\phi'(s')]\int{\cal D}[c]W(\{c\}) \text{e}^{-{\cal
E}_0[\{\phi(s)\}]-{\cal E}_0[\{\phi'(s)\}]}\right]
\nonumber \\
&=&{G\over {\cal Z}_k^0}\text{e}^{-{\cal H}_0}\int_0^Lds' \left[
\int {\cal D}[\phi']\cos[\phi'(s')]\text{e}^{-1/2\int_0^L
ds\left[k'\left(\dot{\phi}'(s)-\dot{\gamma}(s)\right)^2\right]}\right]\nonumber \\
&=&{{\cal Z}_\alpha{\cal Z}_k^0\over {\cal Z}_{\cal
H}^0}\text{e}^{-{\cal H}_0}\int_0^Lds' \left[ {\int {\cal
D}[\phi']\cos[\phi'(s')]\text{e}^{-1/2\int_0^L
ds\left[k'\left(\dot{\phi}'(s)-\dot{\gamma}(s)\right)^2\right]}\over
\int {\cal D}[\phi']\text{e}^{-1/2\int_0^L
dsk'[\dot{\phi}'(s)-\dot{\gamma}(s)]^2}}\right]\nonumber \\
&=&{{\cal Z}_\alpha{\cal Z}_k^0\over {\cal Z}_{\cal
H}^0}\text{e}^{-{\cal H}_0}\int_0^Lds' \left[
\text{e}^{-A'(s',0)/2} \cos[\gamma(s')+\phi_0]\right], \label{R2}
\end{eqnarray}
where
\begin{eqnarray}
k'(s)={k (\alpha+k)\over \alpha+2k},\text{ } \dot{\gamma}(s)={k
\dot{\phi}+\alpha\bar{c}\over \alpha+k},\text{ }\gamma(s)=\int_0^s
\dot{\gamma}(s)ds,\text{ }G=\int{\cal D}[c]\text{e}^{-1/2\int_0^L
ds(2k+\alpha)c^2},\text{ } A'(s,s_0)=\int_{s_0}^s ds/k'(s),
\end{eqnarray}
and we have used an expression similar to Eq. (\ref{partition3})
in the last two lines in Eq. (\ref{R2}) to transform the path
integral into simple integral. $\gamma(s)$ is in general dependent
on $\phi(s)$ and it makes the expression complex. Now the Eq.
(\ref{B21A}) can be reduced into
\begin{eqnarray}
B_2 &\approx&{f\over {\cal Z}_{\cal H}^0}\int {\cal
D}[\phi]B[\{\phi(s)\}]\text{e}^{-{\cal H}_0}\left[ \int_0^Lds'
\text{e}^{-A'(s',0)/2} \cos[\gamma(s')+\phi_0]\right]\nonumber
\\ &=& f\int_0^Lds'\text{e}^{-A'(s',0)/2}\left< B[\{\phi(s)\}]
\cos[\gamma(s')+\phi_0]\right>_\kappa. \label{B23}
\end{eqnarray}
where $\phi_0=\phi(0)$. Eqs. (\ref{B11A}) and (\ref{B23}) are
exactly the same as Eqs. (\ref{B11}) and (\ref{B21}).

\section*{Appendix 4: Calculations of the extension}
In this case $B[\{\phi(s)\}]$ corresponds to cos$\phi(s)$, and so
from Eqs. (\ref{B11})-(\ref{B21}), we obtain
\begin{eqnarray}
X&\equiv&\left<x\right>-\left<x\right>_{f=0}
=\int_0^Lds\left<\cos\phi(s)\right>-\int_0^Lds\left<\cos\phi(s)\right>_\kappa
\nonumber \\
&=&f \int_0^Lds \int_0^Lds'\left[\left< \cos\phi(s)\cos\phi(s')
\right>_\kappa -\text{e}^{-s'/2k'}\left< \cos[\phi(s)]
\cos[\gamma(s')+\phi_0]\right>_\kappa \right]. \label{xA1}
\end{eqnarray}

Note that Eq. (\ref{partition3}) is valid only if $s>s_0$.
Therefore, the integral for $s'$ in Eq. (\ref{xA1}) should be
divided into two parts, one is from $0$ to $s$ and the other is
from $s$ to $L$. When $s>s'$, we have
\begin{eqnarray}
&&\left< \cos\phi(s)\cos\phi(s') \right>_\kappa
=\int_{-\infty}^{\infty}d\phi d\phi'
P(\phi,s;\phi',s')\cos\phi\cos\phi'P(\phi',s';\phi_0,0)
\nonumber\\&=& \text{e}^{-(s-s')/2\kappa}\int_{-\infty}^{\infty}
d\phi' \cos^2\phi'P(\phi',s';\phi_0,0) ={1\over
2}\text{e}^{-(s-s')/2\kappa}\left[1+\text{e}^{-2s'/\kappa}\cos(2\phi_0)\right],
\end{eqnarray}
and
\begin{eqnarray}
&&\left< \cos\phi(s)\cos[\gamma(s')+\phi_0] \right>_\kappa
=\int_{-\infty}^{\infty}d\phi d\phi'
P(\phi,s;\phi',s')\cos\phi\cos\left[\kappa
(\phi'-\phi_0)/\alpha+\phi_0\right]P(\phi',s';\phi_0,0)\nonumber \\
&=&\text{e}^{-(s-s')/2\kappa}\int_{-\infty}^{\infty}d\phi'
\cos\phi'\cos\left[\kappa (\phi'-\phi_0)/\alpha+\phi_0\right]
P(\phi',s';\phi_0,0) \nonumber \\&=&{1\over
2}\text{e}^{-(s-s')/2\kappa-(\alpha+\kappa)^2s'/2\kappa\alpha^2}
\left[\cos(2\phi_0)+\text{e}^{2s'/\alpha}\right].
\end{eqnarray}

When $s<s'$, we obtain
\begin{eqnarray}
&&\left< \cos\phi(s)\cos\phi(s') \right>_\kappa
=\int_{-\infty}^{\infty}d\phi d\phi'
P(\phi',s';\phi,s)\cos\phi\cos\phi'P(\phi,s;\phi_0,0)\nonumber \\
&=&\text{e}^{-(s'-s)/2\kappa}\int_{-\infty}^{\infty}d\phi
\cos^2\phi P(\phi,s;\phi_0,0)={1\over
2}\text{e}^{-(s'-s)/2\kappa}\left[1+\text{e}^{-2s/\kappa}\cos(2\phi_0)\right],\\
&&\left< \cos\phi(s)\cos[\gamma(s')+\phi_0] \right>_\kappa
=\int_{-\infty}^{\infty}d\phi d\phi'
P(\phi',s';\phi,s)\cos\phi\cos\left[\kappa
(\phi'-\phi_0)/\alpha+\phi_0\right]P(\phi,s;\phi_0,0)\nonumber \\
&=&\text{e}^{-\kappa(s'-s)/2\alpha^2}\int_{-\infty}^{\infty}d\phi
\cos\phi\cos[\kappa (\phi-\phi_0)/\alpha+\phi_0]
P(\phi,s;\phi_0,0) \nonumber \\&=&{1\over
2}\text{e}^{-\kappa(s'-s)/2\alpha^2-(\alpha+\kappa)^2s/2\kappa\alpha^2}
\left[\cos(2\phi_0)+\text{e}^{2s/\alpha}\right].
\end{eqnarray}
It follows
\begin{eqnarray}
X&=&2\kappa f\left[ L-{k(k+3\alpha)\over
k+\alpha}+{k(k+\alpha)\over k-\alpha} \text{e}^{-L/k}-{4 \kappa
\alpha\over k-\alpha}
\text{e}^{-L/2\kappa}\right]-{\kappa^2\cos(2\phi_0)\text{e}^{-2L/\kappa}f\over
3(2k+\alpha)(3k+\alpha)}{\cal X},\label{xA1}
\\{\cal X}&=&
6k^2\left( \text{e}^{L/k}-1\right)+k\alpha\left(9
\text{e}^{2L/\kappa}-16\text{e}^{3L/2\kappa}+12\text{e}^{L/k}-5\right)
+\alpha^2\left(3
\text{e}^{2L/\kappa}-8\text{e}^{3L/2\kappa}+6\text{e}^{L/k}-1\right).\nonumber
\end{eqnarray}
In the limit $\alpha \rightarrow k$, $X$ is still finite and
\begin{eqnarray}
X(\alpha \rightarrow k)&=&2\kappa f\left[ L-2k+(L+2k)
\text{e}^{-L/k}\right]-{1\over 12}fk^2\cos(2\phi_0) \left(
1-2\text{e}^{-L/k}+2\text{e}^{-3L/k}-\text{e}^{-4L/k}\right).
\end{eqnarray}

On the other hand, in {\em disorder-first} system,
\begin{eqnarray}
X'&\equiv&\left<x\right>'-\left<x\right>'_{f=0}
=\int_0^Lds\left<\cos\phi(s)\right>'-\int_0^L\left<\cos\phi(s)\right>_\kappa
\nonumber \\
&=&f \int_0^Lds \int_0^Lds'\left< \cos\phi(s)\cos\phi(s')
\right>_\kappa
-f\left[\int_0^Lds\left< \cos[\phi(s)]\right>_\kappa \right]^2\nonumber \\
&=&2\kappa f\left[L+\kappa-\kappa
\left(\text{e}^{-L/2\kappa}-2\right)^2\right] +{\kappa^2f\over
3}\left(\text{e}^{-2L/\kappa}
-6\text{e}^{-L/\kappa}+8\text{e}^{-L/2\kappa}-3\right)\cos(2\phi_0).
\end{eqnarray}
When $\alpha \rightarrow \infty$, it is clearly that $X\rightarrow
X'$.

\section*{Appendix 5: Calculations of the orientational correlation function
and the end-to-end Distance} In this case, $B[\{\phi\}]={\bf
t}(s)\cdot {\bf t}(s')=\cos[\phi(s)-\phi(s')]$. From Eqs.
(\ref{B3})-(\ref{gamma1}), we obtain
\begin{eqnarray}
\left<{\bf t}(s)\cdot {\bf t}(s')\right> &\approx& \left<
\cos[\phi(s)-\phi(s')]\right>_\kappa  \nonumber \\ && +
f\int_0^Lds''\left[\left< \cos[\phi(s)-\phi(s')]\cos\phi(s'')
\right>_\kappa -\text{e}^{-s''/2k'}\left< \cos[\phi(s)-\phi(s')]
\cos[\gamma(s'')+\phi_0]\right>_\kappa\right]. \label{ocfA1}
\end{eqnarray}
The first term in above equation is simple, as
\begin{eqnarray}
&&\left< \cos[\phi(s)-\phi(s')]\right>_\kappa
=\int_{-\infty}^{\infty}d\phi d\phi'
P(\phi,s;\phi',s')\cos(\phi-\phi')P(\phi',s';\phi_0,0)=\text{e}^{-(s-s')/2\kappa}.
\label{ocfA2}
\end{eqnarray}
Again, due to that Eq. (\ref{partition3}) is valid only if
$s>s_0$, the integral for $s''$ in Eq. (\ref{ocfA1}) should be
divided into several parts. If $s>s'>s''$, we have
\begin{eqnarray}
&&\left< \cos[\phi(s)-\phi(s')]
\cos[\gamma(s'')+\phi_0]\right>_\kappa \nonumber
\\&=& \int_{-\infty}^{\infty}d\phi d\phi' d\phi''
P(\phi,s;\phi',s')P(\phi',s';\phi'',s'')\cos(\phi-\phi')\cos[\kappa
(\phi''-\phi_0)/\alpha+\phi_0]P(\phi'',s'';\phi_0,0)\nonumber \\
&=&\text{e}^{-(s-s')/2\kappa}\int_{-\infty}^{\infty} d\phi''
\cos[\kappa (\phi''-\phi_0)/\alpha+\phi_0]P(\phi'',s'';\phi_0,0)%\nonumber \\&=&
=\text{e}^{-(s-s')/2\kappa-\kappa s''/2\alpha^2}\cos\phi_0.
\end{eqnarray}
If $s>s''>s'$, we obtain
\begin{eqnarray}
&&\left< \cos[\phi(s)-\phi(s')]
\cos[\gamma(s'')+\phi_0]\right>_\kappa\nonumber
\\&=&\int_{-\infty}^{\infty}d\phi d\phi' d\phi''
P(\phi,s;\phi'',s'')P(\phi'',s'';\phi',s')\cos(\phi-\phi')\cos[\kappa
(\phi''-\phi_0)/\alpha+\phi_0]P(\phi',s';\phi_0,0)\nonumber \\
&=&\text{e}^{-(s-s'')/2\kappa}\int_{-\infty}^{\infty}d\phi'
d\phi'' P(\phi'',s'';\phi',s')\cos(\phi'-\phi'')\cos[\kappa
(\phi''-\phi_0)/\alpha+\phi_0]P(\phi',s';\phi_0,0)
\nonumber \\
&=&{1\over
2}\text{e}^{-(s-s'')/2\kappa-(\alpha+\kappa)^2(s''-s')/2\kappa\alpha^2}\left(
1+\text{e}^{2(s''-s')/\alpha}\right)\int_{-\infty}^{\infty}d\phi'
\cos[\kappa
(\phi'-\phi_0)/\alpha+\phi_0]P(\phi',s';\phi_0,0)\nonumber \\
&=&{1\over 2}\text{e}^{-s/2\kappa+(3k+\alpha)s'/2k\alpha
-(3k+2\alpha) s''/[2\alpha (k+\alpha)]}\left(
1+\text{e}^{2(s''-s')/\alpha}\right)\cos\phi_0.
\end{eqnarray}
If $s''>s>s'$, we find
\begin{eqnarray}
&&\left< \cos[\phi(s)-\phi(s')]
\cos[\gamma(s'')+\phi_0]\right>_\kappa\nonumber \\
&=&\int_{-\infty}^{\infty}d\phi d\phi' d\phi''
P(\phi'',s'';\phi,s)P(\phi,s;\phi',s')\cos(\phi-\phi')\cos[\kappa
(\phi''-\phi_0)/\alpha+\phi_0]P(\phi',s';\phi_0,0)\nonumber \\
&=&\text{e}^{-\kappa (s''-s)/2\alpha^2}
\int_{-\infty}^{\infty}d\phi d\phi' d
P(\phi,s;\phi',s')\cos(\phi-\phi')\cos[\kappa
(\phi-\phi_0)/\alpha+\phi_0]P(\phi',s';\phi_0,0)\nonumber \\
&=&{1\over 2}\text{e}^{-\kappa
(s''-s)/2\alpha^2-(\alpha+\kappa)^2(s-s')/2\kappa\alpha^2}\left(
1+\text{e}^{2(s-s')/\alpha}\right) \int_{-\infty}^{\infty}d\phi'
\cos[\kappa
(\phi'-\phi_0)/\alpha+\phi_0]P(\phi',s';\phi_0,0)\nonumber \\
&=&{1\over 2}\text{e}^{-(3k+\alpha)(s-s')/2k\alpha-\kappa
s''/2\alpha^2}\left( 1+\text{e}^{2(s-s')/\alpha}\right)\cos\phi_0.
\end{eqnarray}

Similarly, if $s>s'>s''$, then
\begin{eqnarray}
\left< \cos[\phi(s)-\phi(s')]
\cos\left(\phi(s'')\right)\right>_\kappa=\text{e}^{-(s-s')/2\kappa-
s''/2\kappa}\cos\phi_0.
\end{eqnarray}
If $s>s''>s'$, then
\begin{eqnarray}
&&\left< \cos[\phi(s)-\phi(s')]
\cos\left(\phi(s'')\right)\right>_\kappa ={1\over
2}\text{e}^{-(s-3s'+3s'')/2\kappa}\left(
1+\text{e}^{2(s''-s')/\kappa}\right)\cos\phi_0.
\end{eqnarray}
If $s''>s>s'$, then
\begin{eqnarray}
&&\left< \cos[\phi(s)-\phi(s')]
\cos\left(\phi(s'')\right)\right>_\kappa ={1\over
2}\text{e}^{-(3s-3s'+s'')/2\kappa}\left(
1+\text{e}^{2(s-s')/\kappa}\right)\cos\phi_0.\label{ocfA2}
\end{eqnarray}
Combining Eqs. (\ref{ocfA1})-(\ref{ocfA2}), we finally obtain for
$s>s'$
\begin{eqnarray}
\left<{\bf t}(s)\cdot {\bf t}(s')\right> &\approx&
\text{e}^{-(s-s')/2\kappa}+{\kappa \cos\phi_0 f\over
3(3k+\alpha)(k-\alpha)} {\cal S},\label{ocf2p} \\ {\cal S}&=&-6k
(k-\alpha)\text{e}^{-2[(\alpha+2k)s+(\alpha +3k)s']/2k\alpha}
+8\alpha(2k+\alpha)\text{e}^{-s/2\kappa} -6k
(3k+\alpha)\text{e}^{-[2\alpha
s+(k-\alpha)s']/2k\alpha}\nonumber \\
&&+(k-\alpha) (3k+\alpha)\left[ 3\text{e}^{-[(\alpha+k)L+(\alpha
+3k)(s-s')]/2k\alpha}
-3\text{e}^{-[L+3(s-s')]/2\kappa}+6\text{e}^{-s'/2\kappa}
-3\text{e}^{-(L+s'-s)/2\kappa} \right.\nonumber \\
&& \left.
+2\text{e}^{-(4s-3s')/2\kappa}+3\text{e}^{-[(\alpha+k)L+(\alpha
-k)(s-s')]/2k\alpha}\right] .\nonumber
\end{eqnarray}
$\left<{\bf t}(s)\cdot {\bf t}(s')\right>$ is also finite when
$\alpha \rightarrow k$ because in this case,
\begin{eqnarray}
\left<{\bf t}(s)\cdot {\bf t}(s')\right> &\approx&
\text{e}^{-(s-s')/2\kappa}+\kappa \cos\phi_0 f
\left[-\text{e}^{-(L+s'-s)/k} +2\text{e}^{-s'/k}-{13\over
6}\text{e}^{-s/k}+\text{e}^{-L/k}-{1\over
2}\text{e}^{(-3s+2s')/k}\right. \nonumber \\
&&\left.+\text{e}^{-[L+2(s-s')]/k}+{2\over 3}
\text{e}^{(-4s+3s')/k} +\text{e}^{-[L+3(s-s')]/k}+{s'-s\over
k}\text{e}^{-s/k}\right].\label{ocf3}
\end{eqnarray}

In contrast,
\begin{eqnarray}
\left<{\bf t}(s)\cdot {\bf t}(s')\right>' &\approx& \left<
\cos[\phi(s)-\phi(s')]\right>_\kappa +f\int_0^Lds''\left<
\cos[\phi(s)-\phi(s')]\cos\phi(s'') \right>_\kappa \nonumber \\&&
-f\left< \cos[\phi(s)-\phi(s')]\right>_\kappa \int_0^Lds\left<
\cos[\phi(s)]\right>_\kappa \nonumber \\
&=&\text{e}^{-(s-s')/2\kappa} +f\int_0^Lds''\left<
\cos[\phi(s)-\phi(s')]\cos\phi(s'') \right>_\kappa -2\kappa
f\cos\phi_0\text{e}^{-(s-s')/2\kappa}\left(1-\text{e}^{-L/2\kappa}\right)\nonumber \\
&=&\text{e}^{-(s-s')/2\kappa} +{1\over 3}\kappa
f\cos\phi_0\left(6\text{e}^{-s'/2\kappa}-8\text{e}^{-s/2\kappa}
+6\text{e}^{-(L+s-s')/2\kappa}+2\text{e}^{(3s'-4s)/2\kappa}\right.
\nonumber
\\&&\left. -3\text{e}^{-[L+3(s-s')]/2\kappa}-3
\text{e}^{-(L+s'-s)/2\kappa}\right). \label{ocfp}
\end{eqnarray}
When $\alpha \rightarrow \infty$, we obtain $\left<{\bf t}(s)\cdot
{\bf t}(s')\right>\rightarrow\left<{\bf t}(s)\cdot {\bf
t}(s')\right>'$.

The end-to-end distance can be found by
\begin{eqnarray}
R^2&=&\int_0^Lds\int_0^Lds'\left<{\bf t}(s)\cdot {\bf
t}'(s')\right>=2\int_0^Lds\int_0^sds'\left<\cos[\phi(s)-\phi(s')]\right> \nonumber \\
&=&4\kappa L \left[ 1-{2\kappa \over L}\left(1-\text{e}^{-{L\over
2\kappa}}\right)\right]+{4\kappa^2 \cos\phi_0 f \over
9(k-\alpha)^2(k+\alpha)(2k+\alpha)(3k+\alpha)^3}{\cal
Y},\label{endToEnd1A} \\
{\cal Y}&=&18(k-\alpha)^2(k+\alpha)(2k+\alpha)(3k+\alpha)^2L
-9k(k-\alpha)^2(3k+\alpha)^2(4k^2+11\alpha^2+22k\alpha)\nonumber
\\ &&-16\alpha (2k+\alpha )^2[3(k-\alpha)(k+\alpha)(3k+\alpha)L
+4k\alpha (5k+\alpha)(3k-\alpha)]\text{e}^{-L/2\kappa}+18k \alpha
(k+\alpha)^3(k-\alpha)^2\text{e}^{-(1/k+2/\alpha)L}\nonumber
\\ && +18k
(k+\alpha)^3(2k+\alpha)(3k+\alpha)^2\text{e}^{-L/k}-k\alpha
(k-\alpha)^2(2k+\alpha)(3k+\alpha)^2\text{e}^{-2L/\kappa}.
\end{eqnarray}
when $\alpha \rightarrow k$,
\begin{eqnarray}
R^2&=&2k L \left[ 1-{k \over L}\left(1-\text{e}^{-{L\over
k}}\right)\right]+{k\cos\phi_0 f\over
18}\left[36kL-111k^2+(18L^2+78kL+109k^2)\text{e}^{-L/k}+3k^3\text{e}^{-3L/k}
-k^2\text{e}^{-4L/k}\right].\nonumber \\
\end{eqnarray}

On the other hand
\begin{eqnarray}
R'^2&=&4\kappa L \left[ 1-{2\kappa \over
L}\left(1-\text{e}^{-{L\over 2\kappa}}\right)\right] +{4\over
9}\kappa^2 \cos\phi_0 f \left[
18L-99\kappa+16(3L+4\kappa)\text{e}^{-L/2\kappa}+36\kappa\text{e}^{-L/\kappa}
-\kappa\text{e}^{-2L/\kappa}\right].
\end{eqnarray}
When $\alpha \rightarrow \infty$, we obtain $R^2\rightarrow R'^2$.
\end{widetext}

\section* {Acknowledgments}
This work has been supported by the National Science Council of
the Republic of China, National Center for Theoretical Sciences at
Taipei, ROC, and the Natural Sciences and Engineering Research
Council of Canada.

\end{document}